\documentclass[12pt]{article}
\usepackage{pifont}
\usepackage{mathrsfs}
\usepackage{amsfonts}
\usepackage{color}
\usepackage{footmisc}
\usepackage[numbers,sort&compress]{natbib}
\usepackage[bookmarksopen=true, bookmarksopenlevel=\maxdimen,
bookmarksnumbered=true,colorlinks,linkcolor=blue,
citecolor=blue,urlcolor=blue]{hyperref}

\newcommand{\omits}[1]{}

\newcommand{\vect}[1]{\mbox{\boldmath $#1$}}

\definecolor{dyellow}{rgb}{1.,0.8,.0}
\definecolor{myblue}{rgb}{.1,.1,.7}
\definecolor{dcyan}{rgb}{.0,.6,.6}
\definecolor{dmagenta}{rgb}{0.6,0.0,0.6}
\definecolor{brown}{rgb}{0.6,0.2,0.}
\definecolor{darkblue}{rgb}{.0,.0,0.5}
\definecolor{darkred}{rgb}{0.75,0.0,0.0}
\definecolor{orange}{rgb}{1.,.6,.0}
\definecolor{dorange}{rgb}{0.8,.4,.0}
\definecolor{darkgreen}{rgb}{0.0,0.6,0.0}
\definecolor{purple}{rgb}{.4,.0,.4}
\definecolor{lightgrey}{rgb}{0.7,0.7,0.7}

\def\delete{\color{lightgrey}}

\begin{document}
\phantomsection \addcontentsline{toc}{chapter}{Kaluza-Klein-type models of de Sitter and
Poincar\'e gauge theories of gravity}
\begin{center}
{\bf \LARGE Kaluza--Klein-type models of \\ \bigskip de Sitter and
Poincar\'e gauge\\ \bigskip theories of gravity}

\bigskip

\bigskip

{\Large Jia-An Lu$^{a,b}$\footnote{Email: ljagdgz@163.com} and
Chao-Guang Huang$^{b}$\footnote{Email: huangcg@ihep.ac.cn}}

\bigskip

$^a$ School of Mathematics and Computational Science, Sun Yat-sen
University, Guangzhou 510275, China\\

$^b$ Institute of High Energy Physics, and
Theoretical Physics Center for \\
Science Facilities, Chinese Academy of Sciences, Beijing 100049,
China

\phantomsection \addcontentsline{toc}{section}{Abstract}
\begin{abstract}
We construct Kaluza--Klein-type models with a de Sitter or Minkowski
bundle in the de Sitter or Poincar\'e gauge theory of gravity,
respectively. A manifestly gauge-invariant formalism has been given.
The gravitational dynamics is constructed by the geometry of the de
Sitter or Minkowski bundle and a global section which plays an
important role in the gauge-invariant formalism. Unlike the old
Kaluza--Klein-type models of gauge theory of gravity, a suitable
cosmological term can be obtained in the Lagrangian of our models
and the models in the spin-current-free and torsion-free limit
will come back to general relativity with a corresponding
cosmological term. We also generalize the results to the case with a
variable cosmological term.
\end{abstract}
\end{center}

\quad {\small PACS numbers: 04.50.Kd, 04.50.Cd, 02.40.-k}

\section{Introduction}

The first Kaluza--Klein-type model of the gauge theory of gravity is
presented by Mansouri and Chang \cite{Mansouri}, which is similar to
the Kaluza--Klein-type unifications of non-Abelian gauge theories
with gravitation (for example, see Ref. \cite{Cho}). A Kaluza--Klein-type
model of gauge theory of gravity is a gravitational model with
the Lagrangian constructed from the scalar curvature of a fiber
bundle, in which the structure group is the gauge group for gravity
and the Ehresmann connection is related to the geometry of the
spacetime as the base manifold. In the model of Ref.
\cite{Mansouri}, the fiber bundle is the principal bundle with both
the structure group and fiber being the Lorentz or Poincar\'e group,
and is assumed to be torsion free. The parallel transport of vector
fields in the spacetime is used to uniquely relate the gauge
potential in the fiber bundle to the connection of the spacetime.
The scalar curvature of the bundle is equal to the sum of the scalar
curvature of the spacetime, a Yang--Mills Lagrangian and the scalar
curvature of the group space. The action of the model is the
integration of the Lagrangian over the bundle. The model has been
generalized to the torsional case \cite{Guo79}, with the help of
Guo's definition of the torsion field of fiber bundle in
terms of the torsion field of the spacetime \cite{Guo78}.

In these models, however, even if the gauge group is chosen to be
the Poincar\'e group, a cosmological term still appears from the
scalar curvature of the group space. It will prevent the Minkowski
space from being a vacuum solution, or the cosmological term should
be canceled out by hand. How to deal with this problem consistently?
Guo and Chang \cite{Guo79} have proposed the idea of using the
associated Minkowski bundle to solve the problem. In this case, the
cosmological term is replaced by the scalar curvature of the
Minkowski space, which is equal to zero.

In the gauge theories of gravity, the gauge group is usually chosen
to be the Poincar\'e, de Sitter (dS) or anti-de Sitter (AdS) group.
There are several methods to get the gauge-invariant expressions of
the metric and torsion fields. For dS and AdS gauge theory of
gravity established on an umbilical manifold
\cite{{Wu74},{An},{Guo07}}, Guo \cite{Guo76} writes down a dS/AdS-invariant
metric field by making use of the normal vector field of
the manifold. An AdS-invariant metric field is also given in another
geometric framework \cite{Stelle}, in which a global section of the
AdS bundle is used. Locally, the global section of the AdS bundle
corresponds to a 5-vector-valued, non-dynamical scalar field. The
scalar field has further been generalized to the dynamical case
\cite{Stelle}, which is equivalent to using the AdS bundle where the
radius of the AdS fiber is variable. Following Ref. \cite{Stelle}, a
Poincar\'e-invariant metric field has been implicitly used in Ref.
\cite{Grignani}. The explicit expressions for the Poincar\'e-invariant
metric and torsion fields can be found in Ref.
\cite{Salgado}. But, to our knowledge, a gauge-invariant expression
of the torsion field for the case with the dynamical scalar field is still
absent in the literatures.

The gauge-invariant expressions of metric and torsion fields give a
relation between the Ehresmann connection of the bundle and the
geometry of the spacetime. The formalism with such gauge-invariant
expressions will be called the manifestly gauge-invariant formalism.
In this formalism, the configuration variables are the Ehresmann
connection of the principal bundle and a global section of the
associated bundle. The Ehresmann connection is different from the
nonlinear connection used in the nonlinear realization
\cite{Tseytlin}. Generally, the nonlinear connection is related to
the Ehresmann connection in a nonlinear way. When the Ehresmann
connection performs a Poincar\'e, dS or AdS transformation, the
nonlinear connection only performs a Lorentz transformation.

One of the purposes of the present paper is to construct new
Kaluza--Klein-type models with a dS or Minkowski bundle in the dS or
Poincar\'e gauge theory of gravity, respectively. We will define
both the metric and torsion fields on the dS or Minkowski bundle and
calculate the corresponding Riemann--Cartan scalar curvatures. In the
torsion-free case, the Riemann--Cartan scalar curvatures reduce to
the Riemann scalar curvatures in Ref. \cite{Percacci}, where the
scalar curvatures of fiber bundles with generic homogenous fibers in
the framework of Kaluza--Klein theory have been systematically
computed, but the torsional case and the relation between the gauge
potential and the geometry of the spacetime are not taken into
account. The Lagrangian is constructed in such a way that a suitable
cosmological term can be obtained in the model so that the dS or
Minkowski fiber is one of the vacuum solutions of the corresponding
theory. Concretely, in the dS case, the gravitational Lagrangian
consists of two parts: one is the pull back of the scalar curvature
of the dS bundle by a global section and the other is the Lagrangian
for that global section itself. The coupling constant between the
two parts is proportional to the cosmological constant. In the
Poincar\'e case, the gravitational Lagrangian is simply chosen to be
the pull back of the scalar curvature of the Minkowski bundle by a
global section. The global section is just the one used in the gauge-invariant
expressions for the metric and torsion fields. The
gravitational action is the integration of the Lagrangian over some
spacetime region rather than the bundle as in Ref. \cite{Mansouri}.
It will be shown that the scalar curvature of the fiber bundle is a
sum of the scalar curvature of the spacetime, the scalar curvature
of the dS or Minkowski fiber, and a Yang--Mills-like term. The
pullback of the Yang--Mills-like term is merely a quadratic torsion
term. These models will come back to general relativity (GR) with or
without a cosmological term in the spin-current-free and
torsion-free case.

The second purpose of the present paper is to generalize the above
results to the case with a variable cosmological term. The gauge-invariant
expressions for both metric and torsion fields will be
given. For this case the global section used in the gauge-invariant
expressions becomes dynamic and the introduction of its Lagrangian
would become more natural. The variable cosmological term may be of
significance as many variable cosmological constant models could
solve the coincidence problem of the cosmological constant (for
example, see Ref. \cite{Ma}).

The paper is organized as follows. In section 2 we introduce a
manifestly gauge-invariant formalism for the dS and Poincar\'e gauge
theories of gravity. In section 3 we construct concrete Kaluza--Klein-type
models with a dS or Minkowski bundle in the dS or Poincar\'e gauge
theory of gravity, respectively. A suitable cosmological term will
be obtained in these models. The results of sections 2 and 3 are
generalized to the case with a variable cosmological term in section
4. Finally, we end with some remarks in the last section.

\section{de Sitter and Poincar\'e gauge theories of gravity}
We will first introduce the geometric framework for the dS gauge
theory of gravity and then turn to the Poincar\'e case.
In the latter part of the section, we will briefly discuss two specific models of
gauge theory of gravity.

\subsection{de Sitter gauge theory of gravity}

To introduce dS bundle, let ${\cal P}$ be a principal fiber bundle
with the dS group $SO(1,4)$ as its structure group and with the
spacetime manifold ${\cal M}$ as the base space. As $SO(1,4)$ may be
realized at a 5-dimensional (5d) Minkowski space with a fixed
origin, we may set up a 5d Minkowski bundle ${\cal Q}_{{\rm M}_5}$
with a zero section and associate them with ${\cal P}$. A local
section $\sigma$ of ${\cal P}$ presents a local trivialization of
${\cal P}$, which induces a local trivialization on the associated
bundle ${\cal Q}_{{\rm M}_5}$. The local coordinates in the
corresponding region of ${\cal M}$ and Minkowski coordinates in the
typical fiber define the local coordinates on ${\cal Q}_{{\rm
M}_5}$: $\{x^{\mu},~\xi^{A}\}$ ($\mu=0\sim 3,~A=0\sim 4$). The
following condition is gauge invariant and can be used to define a
dS bundle ${\cal Q}_{\rm dS}$:
\begin{equation}\label{dS}
\eta_{AB}\xi^{A}\xi^{B}=l^{2},
\end{equation}
where $l$ is a constant with the dimension of length and the
signature is chosen so that $\eta_{AB}=\textrm{diag}(-1,1,1,1,1)$.
The vertical coordinate basis vector fields
$\partial_{A}=\partial/\partial \xi^{A}$ of ${\cal Q}_{\rm M_5}$
define by projection the vector fields which are tangent to the dS
bundle ${\cal Q}_{\rm dS}$,
\begin{equation}\label{partialA}
\widetilde{\partial}_{A}=\partial_{A}-l^{-2}\xi_{A}\xi^{B}\partial_{B}.
\end{equation}
The horizontal basis vector field of the dS bundle can be chosen as
follows \cite{Percacci}:
\begin{equation}\label{Emu}
E_{\mu}=\partial_{\mu}-\Omega^{A}{}_{B\mu}\xi^{B}\widetilde{\partial}_{A},
\end{equation}
where $\partial_{\mu}=\partial/\partial x^{\mu}$ and
$\Omega^{A}{}_{B\mu}$ is the Ehresmann connection of the principal
bundle in the local section $\sigma$ (cf. Appendix).

Suppose that $\eta_{\cal M}(t)$ is a curve on ${\cal M}$ with
$\eta_{\cal M}(0)=x_0 \in {\cal M}$. Its tangent vector field is
designated by $v^\mu (t) \partial_\mu$. Denote the tangent vector at
$x_0$ by $v_0$. Let $\eta(t)$ be the horizontal lift of $\eta_{\cal
M}(t)$, passing through the point $p_0 \in {\cal Q}_{\rm dS}$ which
lies in the fiber over $x_0$. $\{x^\mu(\eta(t)), \xi^A(\eta(t))\}$
are the coordinates of $\eta(t)$ in the bundle. The tangent vector
field of $\eta(t)$ is required to be $v^\mu (t) E_\mu$. It can be
realized by the following definition of the horizontal lift
\begin{eqnarray} \label{lift}
\frac {dx^\mu(\eta(t))}{dt} =v^\mu(t), \qquad
\frac{d\xi^A(\eta(t))}{dt}= -v^\mu(t) \Omega^A_{\ \, B
\mu}(\eta_\mathcal {M}(t))\xi^B(\eta(t)).
\end{eqnarray}
Then the gauge-covariant derivative of a cross section locally
represented by $\xi^A(x)$ ($x\in {\cal M}$) at $x_0$ with respect to
$v_0$ can
 be defined by
\begin{eqnarray}
D_{v_0}\xi^A(x_0) =\lim_{t\to 0}\frac{\xi^A(\eta_{\cal
M}(t))-\xi^A(\eta(t))}{t},
\end{eqnarray}
where $\xi^A(\eta_{\cal M}(t))$ is the value of the cross section at
$\eta_{\cal M}(t)$. It can be observed that
\begin{eqnarray}\label{derivative0}
D_{v_0}\xi^A(x_0) =[\partial_\mu \xi^A(x_0)+ \Omega^A_{\ \,
B\mu}(x_0)\xi^B(x_0)](v_0)^\mu.
\end{eqnarray}
Remarkably, Eq. (\ref{derivative0}) is easy to be generalized to the
gauge-covariant derivative of a cross section with respect to any
vector field $v$ on ${\cal M}$:
\begin{eqnarray}\label{derivative}
D_{v}\xi^A(x) =[\partial_\mu \xi^A(x)+ \Omega^A_{\ \,
B\mu}(x)\xi^B(x)]v^\mu.
\end{eqnarray}
This derivative will be used in the gauge-invariant expressions of
the metric and torsion fields of the spacetime later.

Now we will show how to tie the bundle structure and the spacetime
structure together. Let ${\cal P}_{H}$ be the right-handed
orthonormal frame bundle of a Riemann--Cartan spacetime manifold
${\cal M}$, where $H=SO(1,3)$ stands for the Lorentz group. Identify
$H$ to a subgroup of $G=SO(1,4)$.  As $H$ act on $G$ by the group
multiplication, we may set up an associated bundle ${\cal P}$ of
${\cal P}_{H}$ with $G$ as the typical fiber. Actually, ${\cal P}$
turns out to be a principal bundle with $G$ as the structure group
\cite{Wise}. Any element of ${\cal P}$ can be expressed by
\begin{equation}
p=p_{H}\cdot g=\{(p_{H}h^{-1},~hg)|~h\in H\},
\end{equation}
where $p_{H}\in {\cal P}_{H}$ and $g\in G$. Suppose that ${\cal M}$
could be covered by finite charts of right-handed orthonormal frame
fields. They corresponds to finite charts of local sections
$\{\sigma_{Hi}(x)\}$ of ${\cal P}_{H}$ and therefore finite charts
of local sections $\{\sigma_{i}(x)=\sigma_{Hi}(x)\cdot I\}$ of
${\cal P}$, where $i$ denotes the $i$th local section and $I$ stands
for the identity element of $G$.  Let
$\mathring{\xi}=(0,0,0,0,l)^{T}$, then $\{\sigma_{i}(x)\cdot
\mathring{\xi}\}$ forms a global section $\phi$ of the dS bundle
${\cal Q}_{\rm dS}$, where
\begin{equation}\label{globalsection}
\sigma_i(x) \cdot \mathring{\xi} = \{(\sigma_i(x)g^{-1}, g \mathring{\xi})| g \in G\}
\end{equation}
is a local section of ${\cal Q}_{\rm M_{5}}$ as well as ${\cal
Q}_{\rm dS}$. In the local section $\sigma_{i}(x)$, the connection
1-form of the principal bundle ${\cal P}$ can be defined as follows:
\begin{equation}\label{connection}
\Omega^{A}{}_{Ba}=\left(
\begin{array}{cc}
\Gamma^{\alpha}{}_{\beta a}&l^{-1}e^{\alpha}{}_{a}\\
-l^{-1}e_{\beta a}&0
\end{array}
\right),
\end{equation}
where $a$ is an abstract index \cite{Wald}, $\alpha, \beta=0\sim3$,
$\{e^{\alpha}{}_{a}\}$ is the dual frame field of the orthonormal
frame field $\{e_{\alpha}{}^{a}\}$, which corresponds to the local
section $\sigma_{Hi}(x)$, and $\Gamma^{\alpha}{}_{\beta a}$ is the
metric-compatible connection 1-form of ${\cal M}$ in
$\{e_{\alpha}{}^{a}\}$. The dual frame fields
$e^{\alpha}{}_{a}$ and the connection 1-forms $\Omega^A{}_{Ba}$,
$\Gamma^{\alpha}{}_{\beta a}$ may also be denoted by
$\vect{e}^{\alpha}$ and $\vect{\Omega}^A{}_B$,
$\vect{\Gamma}^{\alpha}{}_{\beta}$, respectively. The curvature
2-form of $\Omega^{A}{}_{Ba}$, denoted by $\mathcal
{F}^{A}{}_{Bab}$ or $\vect{\cal F}^{A}{}_{B}$, is
\begin{equation} \label{Def_F}
\mathcal
{F}^{A}{}_{Bab}=(d\vect{\Omega}^{A}{}_{B})_{ab}+\Omega^{A}{}_{Ca}\wedge\Omega^{C}{}_{Bb}.
\end{equation}
($d\vect{\Omega}^{A}{}_{B}$ is a 2-form and thus may be denoted as
$(d\vect{\Omega}^{A}{}_{B})_{ab}$ in terms of abstract indices.)
It can be shown that relation (\ref{connection}) is equivalent to
the following manifestly gauge-invariant form:
\begin{equation}\label{metricxi}
\textsf{g}_{ab}=\eta_{AB}(D_{a}\xi^{A})(D_{b}\xi^{B}),
\end{equation}
\begin{equation}\label{torsionxi}
S_{cab}=\mathcal{F}_{ABab}(D_{c}\xi^{A})\xi^{B}.
\end{equation}
They are the metric and torsion field of ${\cal M}$, respectively.
Eq. (\ref{metricxi}) has ever been given in Refs. \cite{Guo76,Stelle},
and the earlier references therein. Here
$\xi^{A}=\xi^{A}(x)$ is the local representation of the global
section $\phi$, $D_{a}\xi^{A}$ is defined by
\begin{equation}\label{Daxi}
v^{a}(D_{a}\xi^{A})=D_{v}\xi^{A}
\end{equation}
for any vector field $v^{a}$ on ${\cal M}$ and can be interpreted as
the local representation of the gauge-covariant derivative of
$\phi$. The metric-compatible connection 1-form
$\Gamma^{\alpha}{}_{\beta a}$ corresponds to a metric-compatible
derivative operator $\nabla_{a}$ of ${\cal M}$ such that
\begin{equation}
\Gamma^{\alpha}{}_{\beta
a}=e^{\alpha}{}_{b}\nabla_{a}e_{\beta}{}^{b}.
\end{equation}
Then the torsion and curvature tensors of ${\cal M}$ are defined as
usual:
\begin{equation}\label{torsion}
(\nabla_{a}\nabla_{b}-\nabla_{b}\nabla_{a})f=
-S^{c}{}_{ab}\nabla_{c}f,
\end{equation}
\begin{equation}\label{curvature}
(\nabla_{a}\nabla_{b}-\nabla_{b}\nabla_{a})\omega_{d}= -R^{c}_{\ \,
dab}\,\omega_{c}-S^{c}{}_{ab}\nabla_{c}\omega_{d}
\end{equation}
for any function $f$ and 1-form $\omega_{a}$ on ${\cal M}$. Let
$S^{\alpha}{}_{ab}=S^{c}{}_{ab}e^{\alpha}{}_{c}$, $R^{\alpha}_{\
\,\beta ab}=R^{c}_{\ \, dab} e^{\alpha}{}_{c}e_{\beta}{}^{d}$, then
Eqs. (\ref{torsion}) and (\ref{curvature}) are equivalent to
\begin{equation}
S^{\alpha}{}_{ab}=(d\vect{e}^{\alpha})_{ab}+\Gamma^{\alpha}{}_{\beta
a}\wedge e^{\beta}{}_{b},
\end{equation}
\begin{equation}
R^{\alpha}_{\ \,\beta
ab}=(d\vect{\Gamma}^{\alpha}{}_{\beta})_{ab}+\Gamma^{\alpha}{}_{\gamma
a}\wedge \Gamma^{\gamma}{}_{\beta b}.
\end{equation}
In fact, similar to the torsion tensor, the curvature tensor also
has the following manifestly invariant form:
\begin{equation}\label{R-F}
R_{cdab}-(2/l^{2})\textsf{g}_{a[c}\textsf{g}_{d]b}=\mathcal
{F}_{ABab}(D_{c}\xi^{A})(D_{d}\xi^{B}).
\end{equation}
Similar to $\Omega^{A}{}_{Ba}$, in the local section
$\sigma_{i}(x)$, $\mathcal {F}^{A}{}_{Bab}$ has the following
expression:
\begin{equation}\label{F}
\mathcal {F}^{A}{}_{Bab}=\left(
\begin{array}{cc}
R^{\alpha}_{\ \, \beta ab}-l^{-2}e^{\alpha}{}_{a}\wedge e_{\beta b}
&l^{-1}S^{\alpha}{}_{ab}\\
-l^{-1}S_{\beta ab}&0
\end{array}
\right).
\end{equation}

According to Eqs. (\ref{metricxi}) and (\ref{torsionxi}), the
gravitational action $S_{\rm G}$ which is a functional of
$\textsf{g}_{ab}$ and $S^{c}{}_{ab}$ could also be viewed as a
functional of $\Omega^{A}{}_{Ba}$ and $\xi^{A}$. It is invariant
under the gauge transformation:
\begin{equation}\label{gauge}
\xi^A\rightarrow g^A{}_B\xi^B,\quad \Omega^A{}_{Ba}\rightarrow
g^A{}_C\Omega^C{}_{Da}(g^{-1})^D{}_B+g^A{}_C\partial_a(g^{-1})^C{}_B,
\end{equation}
where $g^A{}_B=g^A{}_B(x)$ is the matrix representation of a
$G$-valued local function of ${\cal M}$. A $G$-valued local
function of ${\cal M}$ is a smooth map from a region of ${\cal M}$
to the Lie Group $G$. For the de Sitter gauge theory of gravity, the
group $G$ is $SO(1,4)$.

By Eq. (\ref{dS}), there exists $A\in\{0,1,2,3,4\}$, such that
$\xi^{A}\neq 0$ locally. Without loss of generality, it can be
assumed that $\xi^{4}\neq 0$ locally, then $\xi^{4}$ can be viewed
as a function of $\xi^{\alpha}$, according to the condition
(\ref{dS}). The gravitational equations can be given by
\begin{eqnarray}
\delta S_{\rm T}/\delta\Omega=0,\qquad \qquad \qquad \qquad \quad~\label{omega}\\
\delta S_{\rm T}/\delta\xi^{\alpha}+(\delta S_{\rm
T}/\delta\xi^{4})(\partial\xi^{4}/\partial\xi^{\alpha})=0,\label{deltaxi}
\end{eqnarray}
where $S_{\rm T}$ is the total action including the gravitational
action and the action for matter fields. Substituting
$\partial\xi^{4}/\partial\xi^{\alpha}=-\xi_{\alpha}/\xi^{4}$ into
Eq. (\ref{deltaxi}), there will be
\begin{equation}\label{xi}
\delta S_{\rm T}/\delta\xi^{A}=\lambda\xi_{A},
\end{equation}
where $\lambda$ is a local function of spacetime, determined by the
detailed information of the total action.

Actually, Eq. (\ref{xi}) can be deduced from Eq. (\ref{omega}) and
the fact that the action $S_{\rm T}$ is gauge invariant. By Eq.
(\ref{omega}),
\begin{equation} \label{deltaS}
\delta S_{\rm T}=\int(\delta S_{\rm T}/\delta\xi^{A})\delta\xi^{A}.
\end{equation}
We may let
\begin{equation} \label{deltagxi}
\delta\xi^A=\delta[g^A{}_B(x,\lambda)\xi^B(x)]=(\delta
g^A{}_B)\xi^B,
\end{equation}
where $g^{A}{}_{B}(x, \lambda)$ is a family of $SO(1,4)$-valued
local functions, and thus $\delta g^{A}{}_{B}(x)\equiv\frac
\partial {\partial\lambda}|_{\lambda=0}[g^{A}{}_{B}(x, \lambda)]$
is an $\mathfrak{so}(1,4)$-valued local function. Since $S_{\rm T}$
is invariant under the gauge transformation (\ref{gauge}), $\delta
S_{\rm T}$ given by Eqs. (\ref{deltaS}) and (\ref{deltagxi}) should
be equal to zero. Therefore,
\begin{equation}\label{xi2}
(\delta S_{\rm T}/\delta\xi^{A})\xi_{B}-(\delta S_{\rm
T}/\delta\xi^{B})\xi_{A}=0,
\end{equation}
which results in Eq. (\ref{xi}).

\subsection{Poincar\'e gauge theory of gravity}

Now, we turn to the Poincar\'e case. Most of the above formalism is
valid and the differences are as follows. The structure group for
the principal bundle ${\cal P}$ is now the Poincar\'e group
$ISO(1,3)$. The meanings of $\Omega^{A}{}_{Ba}$, $\mathcal
{F}^{A}{}_{Bab}$, $D_{a}$, etc. change to those of the corresponding
objects of the principal Poincar\'e bundle or associated 4d
Minkowski bundle ${\cal Q}_{\rm M_4}$. As a definition of the 4d
Minkowski bundle ${\cal Q}_{\rm M_4}$ from ${\cal Q}_{\rm M_5}$, Eq.
(\ref{dS}) should be replaced by
\begin{equation}\label{xi4}
\xi^{4}=l.
\end{equation}
Eq. (\ref{partialA}) should be replaced by
\begin{equation}\label{partialAP}
\widetilde{\partial}_{\alpha}=\partial_{\alpha},\quad
\widetilde{\partial}_{4}=0 .
\end{equation}
They are tangent to $\cal Q_{\rm M_4}$. Eq. (\ref{connection})
should be replaced by
\begin{equation}\label{omegaP}
\Omega^{A}{}_{Ba}=\left(
\begin{array}{cc}
\Gamma^{\alpha}{}_{\beta a}&l^{-1}e^{\alpha}{}_{a}\\
0&0
\end{array}
\right).
\end{equation}
The gauge-invariant expression for curvature tensor (\ref{R-F})
should be replaced by
\begin{equation}
R_{cdab}=\mathcal {F}_{ABab}(D_{c}\xi^{A})(D_{d}\xi^{B}),
\end{equation}
and Eq. (\ref{F}) replaced by
\begin{equation}\label{FP}
\mathcal {F}^{A}{}_{Bab}=\left(
\begin{array}{cc}
R^{\alpha}_{\ \,\beta ab}
&l^{-1}S^{\alpha}{}_{ab}\\
0&0
\end{array}
\right).
\end{equation}
$g^A{}_B=g^A{}_B(x)$ in Eq. (\ref{gauge}) is now the matrix representation of an
$ISO(1,3)$-valued local function of ${\cal M}$.

The gravitational field equations are given by Eq. (\ref{omega}) and
\begin{equation}\label{Eqxi}
\delta S_{\rm T}/\delta\xi^{\alpha}=0.
\end{equation}
Similar to the dS case, Eq. (\ref{Eqxi}) can be deduced from Eq.
(\ref{omega}) and the gauge-invariant property of the action $S_{\rm
T}$. By Eqs. (\ref{omega}) and (\ref{xi4}),
\begin{equation}
\delta S_{\rm T}=\int(\delta S_{\rm
T}/\delta\xi^{A})\delta\xi^{A}=\int(\delta S_{\rm
T}/\delta\xi^{\alpha})\delta\xi^{\alpha}.
\end{equation}
We may let $\delta\xi^{A}=(\delta g^{A}{}_{B})\xi^{B}$, then $\delta
S_{\rm T}$ given by the above equation will be equal to zero, as
$S_{\rm T}$ is invariant under the gauge transformation
(\ref{gauge}). Now $g^A{}_B=g^{A}{}_{B}(x,\lambda)$ is a family of
$ISO(1,3)$-valued local functions, $\delta g^{A}{}_{B}$ is an
$\mathfrak{iso}(1,3)$-valued local function and therefore,
\begin{equation}
(\delta S_{\rm T}/\delta\xi^{\alpha})\xi_{\beta}-(\delta S_{\rm
T}/\delta\xi^{\beta})\xi_{\alpha}=0,\quad (\delta S_{\rm
T}/\delta\xi^{\alpha})\xi_{4}=0,
\end{equation}
which result in Eq. (\ref{Eqxi}).

\subsection{Example 1: a model of de Sitter gauge theory of gravity}\label{sec:GLW-dSgravity}

Naively, a gauge theory of gravity with a manifest gauge invariance
should have a Yang--Mills-like action for gravitation
\begin{eqnarray}\label{YMAction}
S_{\rm GYM}=-\kappa\int\mathcal {F}^A{}_{Bab}\ \mathcal {F}^B{}_A{}^{ab},
\end{eqnarray}
plus the gauge-invariant action for matter fields, where $\kappa$ is the dimensionless
coupling
constant between matter and gravitational field. For the dS case,
Eq. (\ref{YMAction}) gives \cite{{Wu74},{Townsend},{Guo07}}
\begin{eqnarray}\label{YMActiondS}
S_{\rm GYM}=\kappa\int[R_{abcd}R^{abcd}-\frac{4}{l^{2}}(R-\frac{6}{l^{2}})
+\frac{2}{l^{2}}S_{abc}S^{abc}],
\end{eqnarray}
where Eq. (\ref{F}) is used. The field equations of the dS gravity model are
\begin{eqnarray}\label{1stEq}
\nabla_{c}S_{ab}{}^{c}
+\frac{1}{2}S_{acd}T_{b}{}^{cd}+G_{ba}+\frac 3 {l^2} \textsf{g}_{ab}
=-\frac{l^2}{8\kappa}\Sigma_{ab}
+\frac {l^2} 2(R_{cdea}R^{cde}{}_{b}-\frac {1} 4 R_{cdef}R^{cdef}\textsf{g}_{ab})
\nonumber\\
+S_{cda}S^{cd}{}_{b}-\frac{1}{4}S_{abc}S^{abc}\textsf{g}_{ab},
\end{eqnarray}
\begin{equation}\label{2ndEq}
\nabla_{d}R_{bc}{}^{da} -\frac 1 2T^{a}{}_{de}R_{bc}{}^{de}
+\frac{1}{l^{2}}T^{a}{}_{bc}+\frac{2}{l^{2}}S_{[bc]}{}^{a}=\frac{1}{4\kappa}\tau_{bc}{}^{a},
\end{equation}
where $\Sigma_{ab}$ and $\tau_{bc}{}^a$ are the stress--energy tensor and spin current of
matter fields, respectively,
\begin{displaymath}
G_{ab}=R_{ab}-\frac{1}{2}R\textsf{g}_{ab}, \quad
T^{c}{}_{ab}=S^{c}{}_{ab}+2\delta^{c}{}_{[a}S^{d}{}_{b]d}.
\end{displaymath}

In the torsion-free and vacuum case, the field equations reduce to
\begin{eqnarray}\label{1stEqvac}
G_{ab}+\frac 3 {l^2} \textsf{g}_{ab}
=l^2[C_{acbd} R^{cd}+\frac R 6 (R_{ab}-\frac 1 4 R\textsf{g}_{ab})],
\end{eqnarray}
\begin{equation}\label{2ndEqvac}
\nabla_{d}R^{da}{}_{bc} =0 \qquad \Leftrightarrow\qquad \nabla_{b}R^{a}_{c}
= \nabla_{c}R^{a}_b,
\end{equation}
where $C_{abcd}$ is Weyl curvature tensor. It has been shown that
all the torsion-free vacuum solutions of the dS gravity model are
the vacuum solutions of GR with the same cosmological constant, and
vise versa \cite{{vacuum1},{vacuum2}}.

When the dS gravity model applies to the evolving universe and the matter fields are
composed of spin-current-free, pressureless ideal gas, the
model may explain the accelerating
expansion of the universe and supply a natural transit from
decelerating expansion to accelerating expansion without the help of
dark energy \cite{{cos1},{cos2}}. The torsion together with curvature makes
the universe transit from decelerating expansion to accelerating expansion.
Besides, the attractors in the dS gravity model are analyzed \cite{{AoLiXi},{AoLi}}.

It is remarkable that the dS gravity model becomes highly
non-trivial when the exterior (vacuum) solutions are required to
join some interior solutions \cite{{12},{WFA1},{WFA2}}. For example,
the spin-current-free matter in torsionless interior solutions must
be distributed uniformly \cite{12} because Eq.(\ref{2ndEqvac}) sets
a new constraint.  In torsional case, the exterior solutions can
join with the interior solutions with non-uniformly distributed
matter field, satisfying Newton's law in the weak field approximation,
and supply an alternative way to
 explain the galactic rotation curves without
involving dark matter \cite{{WFA1},{WFA2}}. But unfortunately, the
model may be inconsistent with the solar-system-scale observations,
since the Schwarzschild--dS solutions, which play an important role
in the explanation of the solar-system-scale observations, could not
be smoothly connected to regular internal solutions, in the weak
field approximation \cite{WFA2}.

\subsection{Example 2: a model of Poincar\'e gauge theory of gravity}

For the Poincar\'e case, Eq. (\ref{YMAction}) gives the Stephenson--Kilmister--Yang
action \cite{SKY}
\begin{eqnarray}\label{YMActionP}
S_{\rm GYM}=\kappa\int R_{abcd}R^{abcd},
\end{eqnarray}
where Eq. (\ref{FP}) has been used.  The Einstein term, {\it i.e.},
the scalar curvature term, is absent, and the resulting field
equations are underdetermined at least in the weak field
approximation \cite{WFA1}. In addition, it has been shown
that there is no physical degree in the Stephenson--Kilmister--Yang
action \cite{Maluf}, for the number of constraints is greater than
the number of degrees of freedom in Dirac's prescription for
constrained Hamiltonian systems.

In the following sections, we will
construct several other viable models, by utilizing the
configuration variables, {\it i.e.}, the Ehresmann connection
$\Omega^A{}_{Ba}$ and the global section $\phi$, in the manifestly
gauge-invariant formalism.

\section{Kaluza\texorpdfstring{--}{-}Klein-type models}
In this section, we will construct new Kaluza--Klein-type models of
dS and Poincar\'e gauge theories of gravity with a fixed parameter
$l$.  Again we will discuss the dS case first and turn to the
Poincar\'e case later.

\subsection{New Kaluza\texorpdfstring{--}{-}Klein-type model of dS gauge theory of gravity}

In the dS bundle ${\cal Q_{\rm dS}}$
we may locally define the following 1-form fields
\begin{equation}\label{theta}
\theta^{A}=d\xi^{A}+\Omega^{A}{}_{B\mu}\xi^{B}dx^{\mu},
\end{equation}
which satisfy $\theta^{A}(E_{\mu})=0$. Note that the vector field
$\widetilde{\partial}_{A}$ and the fiber coordinates
$\xi^{A}$ have their corresponding meanings in the
vertical dS spacetime. In the dS spacetime, there exist local
functions $\widetilde{E}_{\alpha}{}^{A}(\xi)$ and
$E^{\alpha}{}_{A}(\xi)$ such that
$\{\widetilde{E}_{\alpha}{}^{A}(\xi)\widetilde{\partial}_{A}\}$ is
an orthonormal frame field with $\{E^{\alpha}{}_{A}(\xi)d\xi^{A}\}$
as its dual frame field. In the dS bundle ${\cal Q}_{\rm dS}$, we
may define
$\widetilde{E}_{\alpha}{}^{A}(x,\xi)=\widetilde{E}_{\alpha}{}^{A}(\xi)$,
$E^{\alpha}{}_{A}(x,\xi)=E^{\alpha}{}_{A}(\xi)$ in a special gauge
first and then let them transform by
\begin{equation}\label{Egauge}
\widetilde{E}_{\alpha}{}^{A}\rightarrow
g^{A}{}_{B}\widetilde{E}_{\alpha}{}^{B},\quad
E^{\alpha}{}_{A}\rightarrow E^{\alpha}{}_{B}(g^{-1})^{B}{}_{A}
\end{equation}
under the local gauge transformation (\ref{gauge}). Let
\begin{equation}
E_{\alpha}=\widetilde{E}_{\alpha}{}^{A}(x,\xi)\widetilde{\partial}_{A},\quad
E^{\alpha}=E^{\alpha}{}_{A}(x,\xi)\theta^{A}.
\end{equation}
Then $\{E_{\mathcal {A}}\}=\{E_{\mu},E_{\alpha}\}$ ($\mathcal
{A}=\mu,~4+\alpha$) becomes a local frame field for ${\cal Q}_{\rm
dS}$ with $\{E^{\mathcal {A}}\}=\{dx^{\mu},E^{\alpha}\}$ as its dual
fame field. Moreover, we may define new functions on the bundle,
$E_{\alpha}{}^{A}$ and $\widetilde{E}_{A}{}^{\alpha}$, by
\begin{equation}
E_{\alpha}{}^{A}=E_{\alpha}(\xi^{A}),\quad
\widetilde{\partial}_{A}=\widetilde{E}_{A}{}^{\alpha}E_{\alpha}.
\end{equation}
They will be used later.

The metric field for ${\cal Q}_{\rm dS}$ can be defined as follows:
\begin{equation}
\overline{\textsf{G}}=\textsf{g}_{\mu\nu}dx^{\mu}\otimes
dx^{\nu}+\eta_{AB}\theta^{A}\otimes\theta^{B}
\end{equation}
with its inverse
\begin{equation}
\overline{\textsf{G}}^{\,-1}=\textsf{g}^{\mu\nu}E_{\mu}\otimes
E_{\nu}+\eta^{AB}\widetilde{\partial}_{A}\otimes\widetilde{\partial}_{B},
\end{equation}
where $\textsf{g}_{\mu\nu}$ is the metric field of the spacetime
manifold, with $\textsf{g}^{\mu\nu}$ as its inverse. Recall that
there is a global section $\phi$ on the dS bundle, defined
by $\{\sigma_i(x)\cdot \mathring{\xi}\}$ and locally represented by $\xi^A(x)$.
In fact, by
Eqs. (\ref{metricxi}) and (\ref{theta}), the pullback
\begin{equation}
\phi^{*}(\textsf{g}_{\mu\nu}dx^{\mu}\otimes
dx^{\nu})=\phi^{*}(\eta_{AB}\theta^{A}\otimes\theta^{B})
\end{equation}
is just the metric field of the spacetime.

The connection coefficients and the curvature components with
respect to $\overline{\textsf{G}}$ in $\{E_{\mathcal {A}}\}$ can be
calculated by the following formulas \cite{Cho}:
\begin{equation} \label{Def_ConnCoef}
\overline{\Gamma}^{\mathcal {C}}{}_{\mathcal {A}\mathcal
{B}}=\frac{1}{2}\overline{\textsf{G}}^{\mathcal {C}\mathcal
{D}}[E_{\mathcal {A}}(\overline{\textsf{G}}_{\mathcal {B}\mathcal
{D}})+E_{\mathcal {B}}(\overline{\textsf{G}}_{\mathcal {A}\mathcal
{D}})-E_{\mathcal {D}}(\overline {\textsf{G}}_{\mathcal {A}\mathcal
{B}})]-\overline{K}^{\mathcal {C}}{}_{\mathcal {A}\mathcal
{B}}-\frac{1}{2}(C^{\mathcal {C}}{}_{\mathcal {A}\mathcal
{B}}+C_{\mathcal {A}\mathcal {B}}{}^{\mathcal {C}}+C_{\mathcal
{B}\mathcal {A}}{}^{\mathcal {C}}),
\end{equation}
\begin{equation}
\overline{R}^{\mathcal{A}}_{\ \ \mathcal{B} \mathcal {C}\mathcal
{D}}=2E_{[\mathcal {C}}(\overline{\Gamma}^{\mathcal
{A}}{}_{|\mathcal {B}|\mathcal {D}]})+2\overline{\Gamma}^{\mathcal
{A}}{}_{\mathcal {E}[\mathcal {C}}\overline{\Gamma}^{\mathcal
{E}}{}_{|\mathcal {B}|\mathcal {D}]}-\overline{\Gamma}^{\mathcal
{A}}{}_{\mathcal{B} \mathcal {E}}C^{\mathcal {E}}{}_{\mathcal
{C}\mathcal {D}},
\end{equation}
where $\overline{K}^{\mathcal {C}}{}_{\mathcal {A}\mathcal {B}}$ is
the contorsion tensor related to the torsion tensor
$\overline{S}^{\mathcal {C}}{}_{\mathcal {A}\mathcal {B}}$ by
\begin{equation}
\overline{K}^{\mathcal {C}}{}_{\mathcal {A}\mathcal
{B}}=\frac{1}{2}(\overline{S}^{\mathcal {C}}{}_{\mathcal {A}\mathcal
{B}}+\overline{S}_{\mathcal {A}\mathcal {B}}{}^{\mathcal
{C}}+\overline{S}_{\mathcal {B}\mathcal {A}}{}^{\mathcal {C}})
\end{equation}
and the structural coefficients $C^{\mathcal {C}}{}_{\mathcal
{A}\mathcal {B}}$ are defined by
\begin{equation}
[E_{\mathcal {A}},E_{\mathcal {B}}]=C^{\mathcal {C}}{}_{\mathcal
{A}\mathcal {B}}E_{\mathcal {C}}.
\end{equation}
After some calculations, the explicit expressions of $C^{\mathcal
{C}}{}_{\mathcal {A}\mathcal {B}}$ can be attained as follows:
\begin{eqnarray}
C^{\sigma}{}_{\mu\nu}=0,~C^{\alpha}{}_{\mu\nu}=-\mathcal
{F}^{A}{}_{B\mu\nu}\xi^{B}\widetilde{E}_{A}{}^{\alpha},\qquad \qquad ~~~\label{C1}\\
C^{\nu}{}_{\mu\alpha}=0,~C^{\beta}{}_{\mu\alpha}=[E_{\mu}(\widetilde{E}_{\alpha}{}^{A})
+\Omega^{A}{}_{B\mu}\widetilde{E}_{\alpha}{}^{B}]\widetilde{E}_{A}{}^{\beta},\label{C2}\\
C^{\mu}{}_{\alpha\beta}=0\qquad \qquad \qquad \qquad \qquad \qquad
\qquad \quad \quad~
\end{eqnarray}
and $C^{\gamma}{}_{\alpha\beta}$ is the same as the structural
coefficients of the orthonormal frame field
$\{\widetilde{E}_{\alpha}^{\ A}(\xi)\widetilde{\partial}_{A}\}$ in
the dS spacetime. Components of the bundle metric field in
$\{E_{\mathcal {A}}\}$ are as follows:
\begin{equation}
\overline{\textsf{G}}_{\mu\nu}=\textsf{g}_{\mu\nu},\quad \overline
{\textsf{G}}_{\mu\alpha}=0,\quad \overline
{\textsf{G}}_{\alpha\beta}=\eta_{\alpha\beta}.
\end{equation}
The bundle torsion can be defined by the following way: the only
nonzero components of $\overline{S}^{\mathcal {C}}{}_{\mathcal
{A}\mathcal {B}}$ are
$\overline{S}^{\sigma}{}_{\mu\nu}=S^{\sigma}{}_{\mu\nu}$
\cite{Guo78}. Then the connection coefficients have the following
expressions:
\begin{eqnarray}
\overline{\Gamma}^{\sigma}{}_{\mu\nu}=\Gamma^{\sigma}{}_{\mu\nu},\quad
\overline{\Gamma}^{\gamma}{}_{\alpha\beta}=\Gamma^{\gamma}{}_{\alpha\beta},\qquad \qquad ~~\\
\overline{\Gamma}^{\alpha}{}_{\mu\nu}=\frac{1}{2}\mathcal
{F}^{A}{}_{B\mu\nu}\xi^{B}\widetilde{E}_{A}{}^{\alpha},\qquad \qquad \qquad ~~~\label{gammabar}\\
\overline{\Gamma}^{\mu}{}_{\nu\alpha}=\overline{\Gamma}^{\mu}{}_{\alpha\nu}=\frac{1}{2}\mathcal
{F}_{AB\nu}{}^{\mu}\xi^{B}E_{\alpha}{}^{A},\qquad \quad ~~\\
\overline{\Gamma}^{\mu}{}_{\alpha\beta}=0,\quad
\overline{\Gamma}^{\beta}{}_{\mu\alpha}=0,\quad
\overline{\Gamma}^{\beta}{}_{\alpha\mu}=-C^{\beta}{}_{\alpha\mu},\label{gammabar2}
\end{eqnarray}
where $\Gamma^{\gamma}{}_{\alpha\beta}$ is the connection
coefficient of the dS metric in the orthonormal frame field
$\{\widetilde{E}_{\alpha}{}^{A}(\xi)\widetilde{\partial}_{A}\}$. To
obtain the above results, some of these formulas are useful:
\begin{eqnarray}
E^{\alpha}{}_{A}E_{\beta}{}^{A}=\delta^{\alpha}{}_{\beta},\quad
\widetilde{E}_{A}{}^{\alpha}\widetilde{E}_{\beta}{}^{A}=\delta^{\alpha}{}_{\beta},\\
E_{\alpha}{}^{A}\widetilde{E}_{B}{}^{\alpha}=\delta^{A}{}_{B}-\xi^{A}\xi_{B}/l^{2},\qquad \quad ~\\
E_{\alpha}{}^{A}=\widetilde{E}_{\alpha}{}^{B}(\delta^{A}{}_{B}-\xi^{A}\xi_{B}/l^{2}),\qquad ~~\\
E_{\alpha}{}^{A}\xi_{A}=0,\quad
\xi^{A}\widetilde{E}_{A}{}^{\alpha}=0,\qquad \qquad
\end{eqnarray}
\begin{equation}\label{C3}
C_{\beta\mu\alpha}=[E_{\mu}(E_{\alpha}{}^{A})
+\Omega^{A}{}_{B\mu}E_{\alpha}{}^{B}]\eta_{AC}E_{\beta}{}^{C}.
\end{equation}
Moreover, it can be shown that the contracted curvature components
are as follows:
\begin{eqnarray}
\overline{R}^{\mu\nu}{}_{\mu\nu}=R_{\mathcal
{M}}-\frac{3}{4}\mathcal
{F}_{AB}{}^{\mu\nu}\mathcal {F}^{A}{}_{C\mu\nu}\xi^{B}\xi^{C},\\
\overline{R}^{\alpha\beta}{}_{\alpha\beta}=R_{F},\qquad \qquad \qquad \qquad \qquad\label{Rbar1}\\
\overline{R}^{\mu\alpha}{}_{\mu\alpha}=\frac{1}{4}\mathcal
{F}_{AB}{}^{\mu\nu}\mathcal {F}^{A}{}_{C\mu\nu}\xi^{B}\xi^{C},
\qquad \quad\label{Rbar2}
\end{eqnarray}
where $R_{\mathcal {M}}$ is the scalar curvature of the spacetime,
$R_{F}$ is the scalar curvature of the typical fiber, $i.e.$, the
scalar curvature of a dS spacetime. Therefore, the scalar curvature
of the dS bundle is
\begin{eqnarray}
\overline{R}&=&\overline{R}^{\mu\nu}{}_{\mu\nu}+
\overline{R}^{\alpha\beta}{}_{\alpha\beta}
+2\overline{R}^{\mu\alpha}{}_{\mu\alpha} \nonumber\\
&=&R_{\mathcal {M}}+R_{F}-(1/4)\mathcal {F}_{AB}{}^{\mu\nu}\mathcal
{F}^{A}{}_{C\mu\nu}\xi^{B}\xi^{C}.\label{Rbar}
\end{eqnarray}

The gravitational Lagrangian may be chosen as
\begin{equation}\label{Lagrangian}
\mathscr{L}_{\rm G}=\chi (\phi^{*}\overline{R}-\frac a {l^2}
\mathscr{L}_{\phi}),
\end{equation}
where
\begin{equation}
\mathscr{L}_{\phi}=\frac 1 2 (D_{a}\xi_{A})D^{a}\xi^{A}+\frac
b {l^2}\xi^{A}\xi_{A}=2+b,
\end{equation}
$\xi^{A}=\xi^{A}(x)$ is the local representation of $\phi$, $\chi$
is the gravitational coupling constant, $a$ and $b$ are two new
dimensionless coupling constants and $0<a<\infty$. In order to
guarantee that the dS space is one of the torsion-free vacuum
solutions of the model, the following condition should hold:
\begin{equation}\label{ab}
a =\frac {18}{2+b} .
\end{equation}
By Eqs. (\ref{F}) and
(\ref{Rbar}),
\begin{equation}\label{phiR}
\phi^{*}\overline{R}=R_{\mathcal {M}}+R_{F}-(1/4)
S_{abc}S^{abc},
\end{equation}
where $R_{F}=4\Lambda =12/l^2$. Therefore, with condition
(\ref{ab}), the Lagrangian (\ref{Lagrangian}) is equal to
\begin{equation}\label{Lagrangian2}
\mathscr{L}_{\rm G}=\chi(R_{\mathcal {M}}-2\Lambda-(1/4)
S_{abc}S^{abc}).
\end{equation}
The gravitational action is the following integration:
\begin{equation}\label{action}
S_{\rm G}=\int_{U}\mathscr{L}_{\rm G}\ \epsilon,
\end{equation}
where $U$ is some spacetime region and $\epsilon$ is the metric-compatible
volume form. This action is dS gauge invariant under the
gauge transformation (\ref{gauge}). The field
equations given by Eqs. (\ref{Lagrangian2}),
(\ref{action}) and
(\ref{omega}) are
\begin{eqnarray}\label{1stEq_KK}
\frac{1}{2}\nabla_{c}S_{ab}{}^{c}
+\frac{1}{4}S_{acd}T_{b}{}^{cd}+G_{ba}+\frac 3 {l^2} \textsf{g}_{ab}=\frac{1}{2\chi}\Sigma_{ab}
+\frac 1 2 S_{cda}S^{cd}{}_{b}-\frac{1}{8}S_{cde}S^{cde}\textsf{g}_{ab},
\end{eqnarray}
\begin{equation}\label{2ndEq_KK}
 T^{a}{}_{bc}+S_{[bc]}{}^{a}=-\frac{1}{\chi}\tau_{bc}{}^{a}.
\end{equation}
Comparison with the dS gravity model in Sec.\ref{sec:GLW-dSgravity}, the
most important features of the theory are that the covariant derivative of the curvature
does not enter Eq. (\ref{2ndEq_KK}) and that Eq. (\ref{2ndEq_KK}) is an algebraic
equation, having the solution,
\begin{equation}
    S_{abc}=-\frac 1 {2\chi}(3\tau_{bca}-\tau_{cab}-\tau_{abc})-\frac 8 {5\chi}
    \textsf{g}_{a[b}\tau_{c]},
\end{equation}
where $\tau_a=\tau_{ac}{}^c$. Therefore, for the gravity coupled to
the matter without spin current, Eq. (\ref{1stEq_KK}) reduces to
the Einstein field equation with the same cosmological
constant. It is remarkable that although the Lagrangian
(\ref{Lagrangian2}) falls into a special case of the general
quadratic models of previous literature\cite{Hayashi}, it is deduced
from first principle here.

\subsection{New Kaluza\texorpdfstring{--}{-}Klein-type models of Poincar\'e gauge theory of gravity}

For the Poincar\'e case, we do not need the concepts of
$\widetilde{E}_{\alpha}{}^{A}$ and $\widetilde{E}_{A}{}^{\alpha}$.
Instead, we may define
\begin{equation}\label{EalphaP}
E_{\alpha}=\partial_{\alpha},\quad E^{\alpha}=\theta^{\alpha}
\end{equation}
in a special gauge and then define
\begin{equation}\label{EalphaAP}
E_{\alpha}{}^{A}=E_{\alpha}(\xi^{A}),\quad
E^{\alpha}{}_{A}=E^{\alpha}(\widetilde{\partial}_{A})
\end{equation}
in an arbitrary gauge, where Poincar\'e version of
$\widetilde{\partial}_{A}$ is given by Eq. (\ref{partialAP}). They
satisfy the following properties:
\begin{eqnarray}
\widetilde{\partial}_{A}=E^{\alpha}{}_{A}E_{\alpha},\quad
\eta_{\alpha\beta}E^{\alpha}{}_{A}\eta^{AB}=E_{\beta}{}^{B},\qquad \qquad ~~\\
E^{\alpha}{}_{A}E_{\beta}{}^{A}=\delta^{\alpha}{}_{\beta},\quad
E_{\alpha}{}^{A}E^{\alpha}{}_{B}=\left\{
\begin{array}{cc}
\delta^{A}{}_{B}\quad A\neq4\\0\quad \quad A=4
\end{array}
\right..
\end{eqnarray}
For the structural coefficients, $C^{\gamma}{}_{\alpha\beta}=0$, the
second formula of Eq. (\ref{C1}) should be replaced by
\begin{equation}
C^{\alpha}{}_{\mu\nu}=-\mathcal
{F}^{A}{}_{B\mu\nu}\xi^{B}E^{\alpha}{}_{A},
\end{equation}
and the second formula of Eq. (\ref{C2}) should be replaced by
\begin{equation}\label{CP}
C^{\beta}{}_{\mu\alpha}=[E_{\mu}(E_{\alpha}{}^{A})
+\Omega^{A}{}_{B\mu}E_{\alpha}{}^{B}]E^{\beta}{}_{A}.
\end{equation}
In the special gauge with respect to Eq. (\ref{EalphaP}), the first
term on the right-hand side of the above equation is equal to zero.
For the connection coefficients, Eq. (\ref{gammabar}) should be
replaced by
\begin{equation}
\overline{\Gamma}^{\alpha}{}_{\mu\nu}=\frac{1}{2}\mathcal
{F}^{A}{}_{B\mu\nu}\xi^{B}E^{\alpha}{}_{A}.
\end{equation}
The gravitational Lagrangian (\ref{Lagrangian}) should be replaced
by
\begin{equation}\label{LagrangianP}
\mathscr{L}_{\rm G}=\chi~\phi^{*}\overline{R}=\chi (R_{\mathcal
{M}}-(1/4)S_{abc}S^{abc}).
\end{equation}
Apparently, {the Largragian and thus the field equations are}
different from the dS case (\ref{Lagrangian2})
only by a cosmological term $\Lambda=3/l^2$. Actually,
they are Poincar\'e gauge
invariant because $\textsf{g}_{ab}$ and $S^a_{\ bc}$ defined by
Eqs. (\ref{metricxi}) and (\ref{torsionxi}) with the gauge group $G=ISO(1,3)$
are now viewed as the Poincar\'e gauge invariants.

\section{Kaluza\texorpdfstring{--}{-}Klein-type models with a variable cosmological term}
As was pointed out by Ref. \cite{Stelle}, we may view $l$ in Eq.
(\ref{dS}) as a positive function of the spacetime. Then
$\partial_{\mu}$ in Eq. (\ref{Emu}) is no longer tangent to the dS
bundle and should be replaced by
\begin{equation}
\check{\partial}_{\mu}=\partial_{\mu}+(\partial_{\mu}l)(\xi^{A}/l)
\partial_{A},
\end{equation}
and thus $E_\mu$ should be replaced by
\begin{equation}\label{Emucheck}
{\check E}_\mu =\check{\partial}_{\mu}-\Omega^A_{\ \, B\mu
}\xi^B\tilde
\partial_{A},
\end{equation}
Accordingly, the second formula of Eq. (\ref{lift}) for the
horizontal lift of $\eta_{\cal M}(t)$ should be replaced by
\begin{equation}
\frac {d\xi^{A}}{dt}=v^{\mu}(t)[(\partial_{\mu}l)(\xi^{A}/l)-\Omega^{A}{}_{B\mu}\xi^{B}],
\end{equation}
and the gauge-covariant derivative (\ref{derivative}) replaced by
\begin{equation}\label{Dxil}
{\check
D}_{v}\xi^{A}(x)=[\partial_{\mu}\xi^{A}(x)-(\partial_{\mu}l)(\xi^{A}/l)
+\Omega^{A}{}_{B\mu}\xi^{B}(x)]v^{\mu}.
\end{equation}
The metric field is given by Eqs.  (\ref{Dxil}), (\ref{metricxi})
and (\ref{Daxi}) with the replacement of $D$ by $\check D$, which is
similar to but slightly different from Eq. (6.6) of Ref.
\cite{Stelle} and has a different explanation. In addition, the
gauge-invariant expression for torsion, Eq. (\ref{torsionxi}),
should be replaced by
\begin{equation}
S_{cab}=\mathcal{F}_{ABab}(\check D_{c}\xi^{A})\xi^{B} -2(
\textsf{g}_{c[a}\partial_{b]}l)/l,
\end{equation}
but the gauge-invariant expression for the curvature tensor is still
given by Eq. (\ref{R-F}) with the replacement of $D$ by $\check D$.
The curvature 2-form is still defined by Eq. (\ref{Def_F}), but in
the local section $\sigma_i(x)$, ${\cal F}^A_{\ Bab}$ in Eq.
(\ref{F}) should be replaced by
\begin{equation}\label{F2}
\mathcal {F}^{A}{}_{Bab}=\left(
\begin{array}{cc}
R^{\alpha}_{\ \,\beta ab}-l^{-2}e^{\alpha}{}_{a}\wedge e_{\beta b}
&l^{-1}S^{\alpha}{}_{ab}+2l^{-2}e^{\alpha}{}_{[a}\partial_{b]}l\\
-l^{-1}S_{\beta ab}-2l^{-2}e_{\beta[a}\partial_{b]}l&0
\end{array}
\right).
\end{equation}
The gravitational equation will be given by Eq. (\ref{omega}) and
\begin{equation}\label{xiEq}
\delta S_{\rm T}/\delta\xi^{A}=0.
\end{equation}

Eq. (\ref{theta}) is now modified to be
\begin{equation}
{\check \theta}^{A}=d\xi^{A}-(\xi^{A}/l)(\partial_{\mu}l)dx^{\mu}
+\Omega^{A}{}_{B\mu}\xi^{B}dx^{\mu},
\end{equation}
so that ${\check \theta}^{A}({\check E}_\mu)=0$. The new definitions
with respect to $\widetilde{E}_{\alpha}{}^{A}$ and
$E^{\alpha}{}_{A}$ are given as follows. In the dS spacetime with
radius $l$, there exist local functions
$\widetilde{E}_{\alpha}{}^{A}(\xi,l)$ and $E^{\alpha}{}_{A}(\xi,l)$
such that
$\{\widetilde{E}_{\alpha}{}^{A}(\xi,l)\widetilde{\partial}_{A}\}$ is
an orthonormal frame field with
$\{E^{\alpha}{}_{A}(\xi,l)d\xi^{A}\}$ as its dual frame field. In
the dS bundle $\check{\cal Q}_{\rm dS}$, we may define
\begin{equation}
\check{\widetilde{E}}_{\alpha}{}^{A}(x,\xi)
=\widetilde{E}_{\alpha}{}^{A}(\xi,l(x)),\quad {\check
E}^{\alpha}{}_{A}(x,\xi)=E^{\alpha}{}_{A}(\xi,l(x))
\end{equation}
in a special gauge and then let them transform by Eq. (\ref{Egauge})
under the gauge transformation (\ref{gauge}). The vertical
orthonormal frame and dual frame fields are
\begin{eqnarray}
{\check E}_\alpha =\check{\widetilde{E}}_{\alpha}{}^{A}(x,\xi))
\widetilde{\partial}_{A}, \qquad {\check E}^\alpha
={\check E}^{\alpha}{}_{A}(x,\xi) {\check \theta}^A.
\end{eqnarray}
The structural coefficients are now defined by
\begin{eqnarray}
[{\check E}_{\cal A}, {\check E}_{\cal B}]  = {\check C}^{\cal C}_{\ \, \cal AB}
{\check E}_C
\end{eqnarray}
with ${\check E}_{\cal A}=\{{\check E}_{\mu}, {\check
E}_{\alpha}\}$. A straightforward calculation shows that the second
formula of Eq. (\ref{C2}) and Eq. (\ref{C3}) should be replaced by
\begin{equation} \label{StruConst-v}
{\check C}^{\beta}{}_{\mu\alpha}=[{\check E}_{\mu} (
\check{\widetilde{E}}_{\alpha}{}^{A}) -(\partial_{\mu}l)(1/l)
\check{\widetilde{E}}_{\alpha}{}^{A} +\Omega^{A}{}_{B\mu}
\check{\widetilde{E}}_{\alpha}{}^{B}]
\check{\widetilde{E}}_{A}{}^{\beta}
\end{equation}
and
\begin{equation}
\check{C}_{\beta\mu\alpha}=[\check{E}_{\mu}(
\check{E}_{\alpha}{}^{A})-(\partial_{\mu}l)(1/l)
\check{E}_{\alpha}{}^{A} +\Omega^{A}{}_{B\mu}
\check{E}_{\alpha}{}^{B}]\eta_{AC}\check{E}_{\beta}{}^{C},
\end{equation}
respectively, with $\check E_\alpha{}^A=\check E_\alpha(\xi^A)$ and
$\check{\widetilde E}_A{}^\alpha$ defined by $\tilde
\partial _A=\check{\widetilde{E}}_{A}{}^{\beta} \check{E}_\beta$. The
connection coefficients, defined by the formula similar to Eq.
(\ref{Def_ConnCoef}) with the replacement of the symbols with
$\check{}$, take the similar results except Eq. (\ref{gammabar2})
should be replaced by
\begin{equation}
\check{\overline{\Gamma}}_{\alpha\mu\beta}=-\check{\overline{\Gamma}}_{\mu\alpha\beta}
=(\partial_{\mu}l)(1/l)\eta_{\alpha\beta},\quad
\check{\overline{\Gamma}}_{\alpha\beta\mu}=-\check{C}_{[\alpha\beta]\mu}.
\end{equation}
For the contracted curvature components, Eqs. (\ref{Rbar1}) and
(\ref{Rbar2}) should be replaced by
\begin{equation}
\check{\overline{R}}{}^{\alpha\beta}{}_{\alpha\beta}=R_{F}-(12/l^{2})
(\nabla_{a}l)\nabla^{a}l
\end{equation}
and
\begin{equation}
\check{\overline{R}}{}^{\mu\alpha}{}_{\mu\alpha}=\frac{1}{4}\mathcal
{F}_{AB}{}^{\mu\nu}\mathcal
{F}^{A}{}_{C\mu\nu}\xi^{B}\xi^{C}-(4/l^{2})(\nabla_{a}l)\nabla^{a}l
-\mathring{\nabla}^{a}[(4/l)\nabla_{a}l]+S^{ab}{}_{a}(4/l)\nabla_{b}l,
\end{equation}
where $\mathring{\nabla}_{a}$ is the torsion-free derivative
operator compatible with $\textsf{g}_{ab}$. As a result, the scalar
curvature (\ref{Rbar}) should be modified to be
\begin{eqnarray}\label{Rbarx}
\check{\overline{R}}=R_{\mathcal {M}}+R_{F}-(1/4)\mathcal
{F}_{AB}{}^{\mu\nu}\mathcal
{F}^{A}{}_{C\mu\nu}\xi^{B}\xi^{C}\qquad \qquad \qquad \nonumber\\
\qquad \qquad \qquad -(20/l^{2})(\nabla_{a}l)\nabla^{a}l
+(8/l)S^{ab}{}_{a}\nabla_{b}l-2\mathring{\nabla}^{a}[(4/l)\nabla_{a}l].
\end{eqnarray}

The gravitational Lagrangian is chosen to be the same as Eq.
(\ref{Lagrangian}), with $\overline{R}$ replaced by
$\check{\overline{R}}$, and $l$ replaced by $l_0$,
where $l_0=$const is a fixed value of $l$-function at some $x$
or a limit value of $l$-function on the spacetime manifold. By Eqs.
(\ref{F2}) and (\ref{Rbarx}),
\begin{eqnarray}\label{phiR2}
\phi^{*}\check{\overline{R}}&=&R_{\mathcal
{M}}+R_{F}-(1/4)S_{abc}S^{abc}\nonumber \\
 &&-(43/2l^{2})(\nabla_{a}l)\nabla^{a}l
+(9/l)S^{ab}{}_{a}\nabla_{b}l-2\mathring{\nabla}^{a}[(4/l)\nabla_{a}l],
\end{eqnarray}
where $R_{F}=4\Lambda=12/l^{2}$ and the last term only contributes
to a boundary term. In the case with $l=l_{0}$,
Eq. (\ref{phiR2})
will come back to Eq. (\ref{phiR}). In order to
guarantee that the dS space with radius $l_0$ is one of the
torsion-free vacuum solutions of the model in the case with $l=l_0$
and $\Lambda_{0}=3/l_{0}{}^{2}$,
the following conditions should hold:
\begin{equation}
\frac a {l_0^2}(2+b)=6\Lambda_0, \ \ ab=-12,
\end{equation}
which result in
\begin{equation}
a=15,\ \ b=-\frac 4 5,
\end{equation}
\begin{equation}
\mathscr{L}_{\phi}=\frac 1 2(D_a\xi_A)D^a\xi^A - \frac 4
5l_0^{-2}\xi^A\xi_A=2 -(4/5)(l/l_0)^2,
\end{equation}
\begin{eqnarray}\label{Lagrangian3}
\mathscr{L}_{\rm G}&=&\chi (\phi^{*} \check{\overline{R}}-5
\Lambda_0\mathscr{L}_{\phi}) \nonumber \\
&=&\chi\{R_{\mathcal{M}}+
6l^{-2}[2-5(l/l_0)^{2}+2(l/l_0)^{4}]-(1/4)S_{abc}S^{abc}
\nonumber \\
&&-(43/2l^{2})(\nabla_{a}l)\nabla^{a}l
+(9/l)S^{ab}{}_{a}\nabla_{b}l-2\mathring{\nabla}^{a}[(4/l)\nabla_{a}l]\}.
\end{eqnarray}
The field equations for the theory are
\begin{eqnarray} \label{torsion-varyl-geq}
&&\frac 1 2 \nabla_c S_{ab}{}^c+\frac 1 4S_a{}^{cd}T_{bcd}
+G_{ba} -3l^{-2}[2-5(l/l_0)^{2}+2(l/l_0)^{4}]\textsf{g}_{ab}  \nonumber \\
&=&\frac 1 {2\chi} \Sigma_{ab}+\frac 1 2 (S^{cd}{}_aS_{cdb}-\frac 1 4 S^{cde}S_{cde}
\textsf{g}_{ab}) \nonumber \\
&&+\frac 9 {2l} [\nabla_a \nabla_b l -\textsf{g}_{ab}\nabla_c\nabla^c l-(\nabla_a l)S^c{}_{bc}]
+\frac {1}{2l^2}[34(\nabla_a l)(\nabla_b l)-\frac {25} 2
\textsf{g}_{ab}(\nabla_c l)(\nabla^c l)], \qquad
\end{eqnarray}
\begin{eqnarray} \label{torsion-varyl-torsioneq}
S_{[bc]}{}^a + T^a{}_{bc}&=& - \frac 1 \chi \tau_{bc}{}^a
+ \frac 9 l \delta^a{}_{[b}\nabla_{c]}l,
\end{eqnarray}
\begin{eqnarray} \label{torsion-varyl-leq}
\frac {43}{l}[\mathring{\nabla}_a\nabla^a l -l^{-1}(\nabla_a l)(\nabla^a l) ]
+24l^2(l_0^{-4}-l^{-4})&=&9 \mathring{\nabla}_a S^{ba}{}_b.
\end{eqnarray}
The last line of Eq. (\ref{torsion-varyl-geq}) can be regarded as the stress--energy
tensor for scalar field $l$.  Again, Eq. (\ref{torsion-varyl-torsioneq}) is an algebraic
equation and has the solution
\begin{equation}
S_{abc}=-\frac 1 {2\chi}(3\tau_{bca}-\tau_{cab}-\tau_{abc})-\frac 8 {5\chi}
    \textsf{g}_{a[b}\tau_{c]}-\frac {18}{5l}\textsf{g}_{a[b}\nabla_{c]}l .
\end{equation}
It means that the variable $l$ also serves as the source of torsion.
Therefore, among the following three conditions, two are satisfied then
the third one must be also satisfied and the gravity reduces to GR
with cosmological constant $\Lambda_0$:
\begin{eqnarray}
  S_{abc} &=& 0, \\
  \tau_{bca} &=& 0, \\
  l & = & l_0.
\end{eqnarray}

For the Poincar\'e case, Eq. (\ref{F2}) should be replaced by
\begin{equation}
\mathcal {F}^{A}{}_{Bab}=\left(
\begin{array}{cc}
R^{\alpha}_{\ \,\beta ab}&l^{-1}S^{\alpha}{}_{ab}+2l^{-2}e^{\alpha}{}_{[a}\partial_{b]}l\\
0&0
\end{array}
\right).
\end{equation}
We do not need the concepts of
$\check{\widetilde{E}}_{\alpha}{}^{A}$ and
$\check{\widetilde{E}}_{A}{}^{\alpha}$. Instead, we may define
$\check E_{\alpha}{}^{A}$ and $\check E^{\alpha}{}_{A}$ by Eqs.
(\ref{EalphaP}) and (\ref{EalphaAP}) with the replacement of the
symbols with $\check{}$. Eq. (\ref{StruConst-v}) should be replaced
by
\begin{equation}
\check{C}^{\beta}{}_{\mu\alpha}=[\check{E}_{\mu}(\check{E}_{\alpha}{}^{A})-(\partial_{\mu}l)(1/l)
\check{E}_{\alpha}{}^{A}+\Omega^{A}{}_{B\mu}
\check{E}_{\alpha}{}^{B}]\check{E}^{\beta}{}_{A},
\end{equation}
and Eq. (\ref{Lagrangian3}) replaced by
\begin{equation}\label{LagrangianPl}
\mathscr{L}_{\rm G}=\chi~\phi^{*}\check{\overline{R}},
\end{equation}
where $\phi^{*}\check{\overline{R}}$ is given by Eq. (\ref{phiR2})
with $R_F=0.$ The second field equation is identical to Eq.
(\ref{torsion-varyl-torsioneq}), and the other two equations are now
\begin{eqnarray}\label{KKlxP-eq1}
&&\frac 1 2\nabla_c S_{ab}{}^c+\frac 1 4 S_a{}^{cd}T_{bcd}+G_{ba}
=\frac 1{2\chi} \Sigma_{ab}+ \frac 1 2\left (S^{cd}{}_a S_{cdb}-
\frac 1 4S_{abc}S^{abc}\textsf{g}_{ab}\right ) \nonumber \\
&&+\frac 9{2l}\left (\nabla_a\nabla_bl-\textsf{g}_{ab}\nabla_c\nabla^c l
-(\nabla_al)S^c{}_{bc}\right ) +\frac 1 {2l^2}[34 (\nabla_a l)\nabla_b l
-\frac {25}2\textsf{g}_{ab}(\nabla_c l)(\nabla^c l)],\qquad
\end{eqnarray}
\begin{equation}\label{KKP-eq3}
l\mathring{\nabla}_a\nabla^a l-(\nabla_a
l)(\nabla^a l)= \frac {9l^2} {43}\mathring{\nabla}_a S^{ba}{}_b.
\end{equation}
The local Poinar\'e gauge-invariant
model can reduce to GR under the
same conditions as the above local dS gauge-invariant model, with $l=l_0$ replaced
by $l =$const. An important difference is that the Minkowski space is a
vacuum solution of the model in those cases.

\section{Remarks}

In these new Kaluza--Klein-type models, we use the Riemann--Cartan
scalar curvature of the associated bundle and a global section to
construct the gravitational dynamics. The action is the integration
of the Lagrangian over some spacetime region as usual. A suitable
cosmological term can be obtained so that the dS or Minkowski space
as the typical fiber is one of the vacuum solutions of the theory.
In the spin-current-free and torsion-free limit, the models reduce to GR with
the same cosmological term. It should be mentioned that these models
are different from the original Kaluza--Klein-type
models\cite{Mansouri,Guo79} of gauge theory of gravity, which use
the Riemann or Riemann--Cartan scalar curvature of a principal
bundle to serve as the gravitational Lagrangian and use the
integration of the Lagrangian over the principal bundle to serve as
the action. In addition, the old models can not reduce to GR in the
spin-current-free and torsion-free limit and have not provided a rule to get
a suitable cosmological term.

It should be emphasized that both dS and Poincar\'e gauge theories
of gravity presented in this paper are manifestly gauge invariant.
In the formalism, the configuration variables are Ehresmann
connection $\Omega^A_{\ Ba}$ and vector-valued scalar field $\xi^A$
which are covariant under the gauge transformations (\ref{gauge}).
The geometric variables, such as metric $\textsf{g}_{ab}$ and
torsion $S^{c}{}_{ab}$, are expressed as the functions of
$\Omega^A_{\ Ba}$ and $\xi^A$ and are invariant under the gauge
transformations. The actions are the functionals of $\Omega^A_{\
Ba}$ and $\xi^A$, which are invariant under the gauge
transformations. The manifestly gauge-invariant formalism is
motivated by the principle of localization \cite{Guo07,Guo08}, which
states that gravity should be based on the localization of the full
symmetry of the corresponding special relativity (SR) as well as
dynamics. It should be noted that the manifestly gauge-invariant
formalism may also be applied to SR and GR. For example, in
Einstein's SR, the flat metric field can be expressed in a local
Poincar\'e gauge-invariant form via Eqs. (\ref{derivative}),
(\ref{metricxi}) and (\ref{omegaP}), while the Ehresmann connection
is constrained by the flat condition $\mathcal {F}^A{}_{Bab}=0$, and
the vector-valued scalar field is constrained by the condition
$\xi^4={\rm const}\neq 0$. In GR, the metric field can be expressed
in a local Poincar\'e gauge-invariant form via the above mentioned
equations, while the Ehresmann connection and the vector-valued
scalar field are constrained by the torsion-free condition $\mathcal
{F}_{ABab}(D_c\xi^A)\xi^B=0$ and the condition $\xi^4={\rm
const}\neq 0$. In contrast, in the Poincar\'e gauge theory of
gravity discussed in section 3 of this paper, the Ehresmann
connection is completely determined by the variation principle,
while the vector-valued scalar field is still constrained by the
condition $\xi^4={\rm const}\neq 0$. In the gauge with
$\xi^{A}=(0,0,0,0,l)^{T}$, it is easy to see that the actions which
are functionals of $\Omega^A_{\ Ba}$ and $\xi^A${\delete ,} can also
be viewed as the functionals of $\Gamma^{\alpha}{}_{\beta a}$ and
$e^{\alpha}{}_{a}$, which can be identified with the nonlinear
connection in the nonlinear realization \cite{Tseytlin}. The
formalism with the nonlinear connection as variable emphasizes on
the Lorentz invariance of the theory, which confirms the remaining
symmetry in the gauge with $\xi^{A}=(0,0,0,0,l)^{T}$.

The actions for both dS and Poincar\'e gauge theories of gravity can
be written as the functionals of metric and torsion fields. It is
remarkable that once the actions in different gauge theories of
gravity are written in these forms, it is difficult to say that they
are dS gauge invariant or Poincar\'e gauge invariant or even only Lorentz
gauge invariant. The information of different gauge groups have been
hidden in the concrete expressions of metric and torsion fields in
terms of $\Omega^A_{\ Ba}$ and $\xi^A$. What can be confirmed is
that such actions are diffeomorphism invariant and Lorentz invariant
through the tetrad formalism. One may try to determine the gauge
group by observing whether the dS spacetime or Minkowski spacetime
is one of the vacuum solutions of the corresponding theory. It is
interesting to see whether there are any observational effects to
distinguish the internal symmetries.

We have also considered the case where the Ehresmann
connection and the vector-valued scalar field are completely
determined by the variation principle. As a result, the global section
becomes dynamic, and a variable cosmological
term appears in the Kaluza--Klein-type Lagrangian. Actually,
many variable cosmological constant models may solve the coincidence
problem of the cosmological constant and numerous works have been
done to search for the theoretical foundation of such models
\cite{Ma}. If our models turn out to be able to consistently explain
the observational data of the universe while solving the coincidence
problem, they would serve as elegant explanations for the
accelerating expanding universe.

Although only the dS and Poincar\'e cases have been discussed in
this paper, it is an easy thing to modify the results of the dS case
to obtain the corresponding results of the AdS case.

\phantomsection \addcontentsline{toc}{section}{Acknowledgments}
\section*{Acknowledgments}
This paper is in memory of Prof. Han-Ying Guo. We are grateful to
him for illuminating discussions. We would like to thank Prof. Zhan
Xu, Prof. Xiao-Ning Wu, Prof. Yu Tian and Prof. Yi Ling for useful
discussions. One of us (Lu) would like to thank the late Prof. Han-Ying
Guo for his guidance. He would also like to thank Prof. Zhi-Bing Li
and Prof. Xi-Ping Zhu for their help and guidance. He also thanks
the hospitality during his stay at Institute of High Energy Physics,
and Theoretical Physics Center for Science Facilities, Chinese
Academy of Science. This work is supported by National Science
Foundation of China under Grant Nos. 10975141, 10831008, 11275207,
and the oriented projects of CAS under Grant No. KJCX2-EW-W01.

\phantomsection \addcontentsline{toc}{section}{Appendix. Ehresmann connection}
\section*{Appendix. Ehresmann connection}

Consider a generic principal fiber bundle ${\cal P}({\cal M}, G)$ and its
tangent bundle $T{\cal P}$.  $T{\cal P}$ may be decomposed as %
\begin{eqnarray}
T{\cal P} = {\cal V}\oplus{\cal H},
\end{eqnarray}
where ${\cal V}$ and ${\cal H}$ are the smooth vertical and horizontal subbundles,
respectively, and $\oplus$ is a direct sum.
The vertical subbundle is defined canonically by the fundamental vector fields,
which are generated by elements in the Lie algebra $\frak{g}$ of the bundle
${\cal P}({\cal M}, G)$ and are tangent to $G_p$ at every point $p\in {\cal P}$.
The horizontal subbundle is an Ehresmann connection in terms of distributions.
\cite{KobayashiNomizu}

If and only if
\begin{equation}
    {\cal H}_{pg} = d(R_g)_p({\cal H}_{p}), \qquad p\in {\cal P}, g \in G,
\end{equation}
the Ehresmann connection defines a Lie algebra valued 1-form
$\vect{\nu} \in \frak{g}\otimes \Omega^1({\cal P})$:
\begin{equation}\label{nu}
    \vect{\nu}: T_p{\cal P} \to {\cal V}_p \simeq {\frak g}, \qquad p\in {\cal P}
\end{equation}
such that ${\cal H}={\rm ker}({\vect{\nu}})$,
\begin{eqnarray}
&& \qquad \vect{\nu}(\vect{\xi}) = \xi, \qquad\qquad\qquad\qquad \
\xi \in {\frak g};  \\
&& \qquad R^*_g\vect{\nu} = {\rm ad}_{g^{-1}}\vect{\nu} = g^{-1}\vect{\nu} g, \qquad g\in G,
\end{eqnarray}
where $R_g$ denotes right multiplication by $g$, $\vect{\xi}$ is the
fundamental vector field on ${\cal P}$ associated with $\xi$ by differentiating the $G$ action on ${\cal P}$, and
ad$_{g}(X):= \left . \frac {d}{dt}g \exp(tX)g^{-1}\right |_{t=0}$ is
the adjoint action.

Locally, on a given coordinate patch $U_i$ of ${\cal M}$ and for a given
local sections $\sigma_i$,
the pullback $\sigma^*_i$ and $\vect{\nu}$ may define a local
connection $\vect{A}_i \in \frak{g} \otimes \Omega^1(U_i)$ by
\begin{equation}
\vect{A}_i \equiv \sigma^*_i \vect{\nu}.
\end{equation}
It is the local form of the Ehresmann connection $\vect{\nu}$ and
may be identified with the gauge potential up to some Lie algebra
factor. Remember that $\vect{\nu}$ is globally defined. In the
intersection $U_i\bigcap U_j$ of two coordinate patches $U_i$ and
$U_j$, the local connection should transform as
\begin{equation}
\vect{A}_j = \Lambda_{ji} \vect{A}_i \Lambda^{-1}_{ji} + \Lambda_{ji} d(\Lambda^{-1}_{ji}),
\end{equation}
where $\Lambda_{ij}$ is the transition function from $U_i$ to $U_j$.
(The repeated indices are not summed up, here.)
Similarly, for the local sections $\sigma_i$ and $\tilde \sigma_i$ in
two given gauges, where $\tilde \sigma_i =\sigma_i (g_i)^{-1}$ with $g_i\in G$, the local
connections $\vect{A}_i$ and $\tilde{\vect{A}}_i$ for the two sections transform as
\begin{equation}\label{gaugetransf}
\tilde{\vect{A}}_i = g_i \vect{A}_i g^{-1}_i + g_i dg^{-1}_i,
\end{equation}
where $i$ is not summed up.

Expanding $\vect{A}$ in terms of Lie algebra
generators $\tau_a$ and the dual coordinate bases $dx^\mu$ we have
\begin{equation}
\vect{A} = \vect{A}^a  \tau_a = A^a{}_\mu \tau_a
dx^\mu,
\end{equation}
where the index for the $i$th coordinate patch has been omitted.
When a matrix representation of the Lie algebra ${\frak g}$ is used,
we have the matrix form of $\vect{A}$:
\begin{equation}
\vect{A}^A{}_B =A^A{}_{B\mu}dx^\mu.
\end{equation}
Then, Eq. (\ref{gaugetransf}) can be written as
\begin{equation}
\tilde{\vect{A}}^A{}_B = g^{A}{}_C \vect{A}^C{}_D (g^{-1})^D{}_B + g^A{}_C d(g^{-1})^C{}_B.
\end{equation}
In the notation of Ref. \cite{Wald}, $\vect{A}^A{}_B$ is also
written as $A^A{}_{Ba}$.

The connection $\Omega^A{}_{B\mu}$ in Eq. (\ref{Emu}) is defined in
the same way as $A^A{}_{B\mu}$. We do not use the notation
$\Omega^A{}_{B\mu}$ in this appendix, for the reason that it is
similar to that of the 1-form space $\Omega^1$, which appears in the line above
Eq. (\ref{nu}).

\phantomsection 
\end{document}